\documentclass[conference]{IEEEtran}
\IEEEoverridecommandlockouts
\usepackage{cite}
\usepackage{amsmath,amssymb,amsfonts}
\usepackage{algorithmic}
\usepackage{graphicx}
\usepackage{textcomp}
\usepackage{xcolor}
\usepackage{soul}
\usepackage{float}
\usepackage{caption}

\def\BibTeX{{\rm B\kern-.05em{\sc i\kern-.025em b}\kern-.08em
    T\kern-.1667em\lower.7ex\hbox{E}\kern-.125emX}}

\begin{document}

\title{A Bit-Parallel Deterministic Stochastic Multiplier }

\author{\IEEEauthorblockN{
Sairam Sri Vatsavai and 
Ishan Thakkar}
\IEEEauthorblockA{Department of Electrical and Computer Engineering,
University of Kentucky, Lexington, KY 40506, USA}}
\vspace{-7em}

\maketitle

\begin{abstract}
 This paper presents a novel bit-parallel deterministic stochastic multiplier, which improves the area-energy-latency product by up to  10.6$\times$10$^4$, while improving the computational error by 32.2\%, compared to three prior stochastic multipliers.      
\end{abstract}

\section{Introduction}
Stochastic Computing (SC) is an unconventional form of computing where numbers are represented by the probability of observing a '1' in bit-streams called stochastic bit-streams (SBs) \cite{Gaines1969}. In SC's unipolar format, \textit{W} is an SB of \textit{N} bits that represents a real-valued variable $\upsilon\in[0,1]$, $\upsilon$= $N_1/N$, where $N_1$ is the number of '1's in \textit{W}. SC offers a low-cost multiplication using a standard logic AND gate  \cite{Gaines1969}. Therefore, stochastic multipliers can decrease the hardware complexity of GEMM circuits used in deep learning accelerators \cite{ugemm}. \textit{\textbf{But}} Stochastic multipliers suffer from computational errors. To reduce errors, prior works \cite{Gaines1969,Jenson2016} used lengthy pseudo-random SBs. In contrast, \cite{ugemm} showed that errors can be minimized by deterministically re-adjusting the bit-position correlations in randomly generated SBs. Atop errors, prior stochastic multipliers also suffer from very high latency and energy due to their use of lengthy SBs and bit-serial operation.  
To address both of these challenges, we present, \textbf{\textit{for the first time}}, a novel stochastic multiplier that generates SBs with reduced lengths and deterministic bit-position correlations in a bit-parallel manner, thereby simultaneously minimizing the latency, energy, and errors.

\section{Our Stochastic Multiplier}
\begin{figure}[H]
\captionsetup{font=footnotesize}
    \centering
    \includegraphics[scale=0.27]{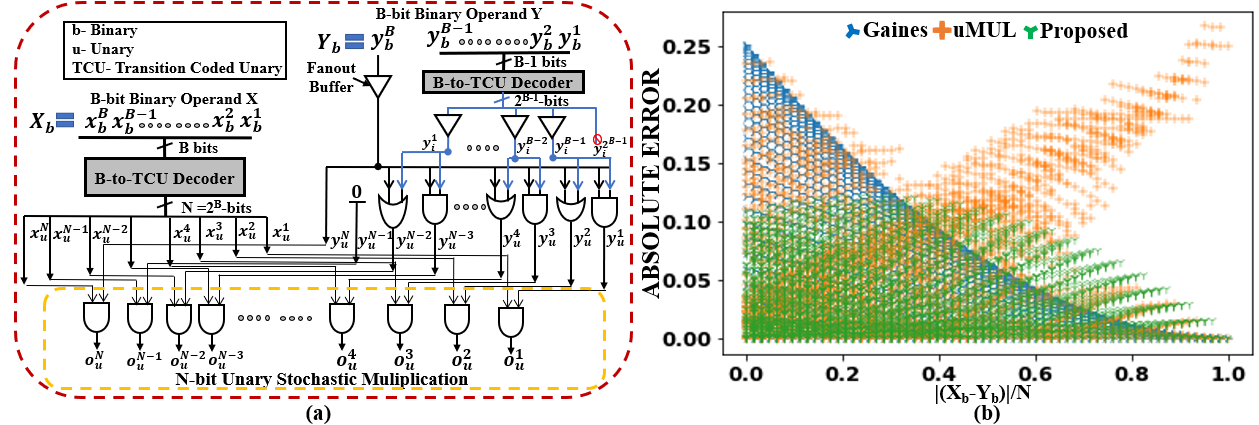}
    \caption{(a) Schematic of our proposed stochastic multiplier, (b) distribution of absolute error in various stochastic multipliers.}
    \label{stochasticmultiplier}
\end{figure}

 Fig. \ref{stochasticmultiplier}(a) shows our stochastic multiplier, which first converts two B-bit binary operands (X$_b$ and Y$_b$) to N-bit SBs ($X_u$=$[x_u^N,..,x_u^1]$ and $Y_u$=$[y_u^N,..,y_u^1]$), where N=2$^B$. Subsequently, it performs bit-wise AND on $X_u$ and $Y_u$ to obtain the stochastic multiplication result $O_u$=$[o_u^N,..,o_u^1]$. To minimize the errors, the conditional probability P($Y_u$/$X_u$) must be equal to the marginal probability P($X_u$) \cite{ugemm}. Our multiplier achieves that as follows. From Fig. \ref{stochasticmultiplier}(a), for operand X$_b$, a binary-to-transition-coded-unary (B-to-TCU) decoder generates all N bits of $X_u$ with '1's grouped at the trailing end (on the right). For operand Y$_b$, only the binary bits $[y_b^{B-1},..,y_b^1]$ of Y$_b$ go to the B-to-TCU decoder to generate bits $[y_i^{2^{B-1}},..,y_i^1]$. These bits together with $y_b^B$ propagate through the bit-position correlation encoder (the array of AND and OR gates) to generate $Y_u$, while maintaining P($Y_u$/$X_u$)=P($X_u$). Once $X_u$ and $Y_u$ are available, they are pushed through the array of AND gates to obtain $O_u$. Table I reports examples of how our proposed design generates $X_u$ and $Y_u$, and then multiplies them using AND gates to generate $O_u$.
 
 %Table I also shows for the bottom two cases how stochastic multiplication output can result in errors. 
 %In example 1, X$_b$=$100$, represented as P(x$_u$=1)=4/8, by the number of '1's in $X_u$. Similarly, to represent Y$_b$=$110$ as P(y$_u$=1)=6/8, our multiplier (Fig. \ref{stochasticmultiplier}(a)) sends \{$y_b^{2},y_b^{1}$=$1 ,0$\} to B-to-TCU decoder to generate \{$y_i^{4},y_i^{3},y_i^{2},y_i^{1}$=$0,0,1,1$\}. Then, \{$y_b^3$=1\} and \{$y_i^{1},y_i^{2},y_i^{3},y_i^{4}$\} pass through the array of AND and OR gates to generate the unary output $Y_u$ representing P(y$_u$=1)=6/8. Then, parallel bit-wise AND is carried out between $X_u$ and $Y_u$, employing the array of AND gates, to generate output $O_u$ that represents P(o$_u$=1)=3/8, which is equivalent to the multiplication of P(x$_u$=1)$\times$P(y$_u$=1). However, stochastic multiplication output can result in errors, as shown for the other two examples in Table I. 
 
 %We quantified the errors and hardware cost of the proposed multiplier in Section \ref{3}.    
\vspace{0.5em}
\begin{table}[H]
\captionsetup{font=scriptsize}
\label{table1}
\centering
\caption*{TABLE I: EXAMPLES FOR OUR STOCHASTIC MULTIPLIER. ERROR IS THE DIFFERENCE BETWEEN THE TARGET AND ACTUAL OUTPUT PROBABILITIES.}

\begin{tabular}{|c|c|c|c|}
\hline
\textbf{$X_u$=P(x$_u$=1)}   & \textbf{$Y_u$=P(y$_u$=1)}   & \textbf{$O_u$=P(o$_u$=1)} & \textbf{Error} \\ \hline
00001111=4/8 & 10111110=6/8 &  00001110=3/8         & 0             \\ \hline
00011111=5/8 & 00101010=3/8 & 00001010=2/8         & 0.01          \\ \hline
00000111=3/8 & 10101010=4/8 & 00000010=1/8         & 0.06           \\ \hline
\end{tabular}
\end{table}
\vspace{0.5em}
\begin{table}[H]
\centering
\captionsetup{font=scriptsize}
\caption*{TABLE II: COMPARISON OF STOCHASTIC MULTIPLIERS. A=AREA, L=LATENCY, E=ENERGY, MAE=MEAN ABSOLUTE ERROR}
\label{table2}
\begin{tabular}{|c|c|c|c|c|c|c|}
\hline
\textbf{Unit}                            & \textbf{\textbf{\begin{tabular}[c]{@{}c@{}}$A$\\ ($\mu$$m^2$)\end{tabular}}} & \textbf{\begin{tabular}[c]{@{}c@{}}$L$\\ (ns)\end{tabular}} &  \textbf{\begin{tabular}[c]{@{}c@{}}$E\times L$\\ (pJ.s)\end{tabular}} & \textbf{\begin{tabular}[c]{@{}c@{}}$A\times E \times L$\\ (pJ.s.mm$^2$)\end{tabular}} & \textbf{MAE} \\ \hline
\textbf{uMUL\cite{ugemm}}                            & 207.6         & 640              & 2.5E-08      & 5.2E-09                    & 0.06         \\ \hline
\textbf{Gaines\cite{Gaines1969}}                          & 378.7         & 640              &  4.9E-08      & 1.9E-08                    & 0.08         \\ \hline
{\color[HTML]{262626} \textbf{Jenson\cite{Jenson2016}}}   & 520.2         & 163840           &  3.5E-03      & 1.8E-03                    & 0.07         \\ \hline
{\color[HTML]{262626} \textbf{Proposed}} & 540.6         & 0.17             &  9.2E-14      & 4.9E-14                    & 0.04         \\ \hline
\end{tabular}
\end{table}

\section{Evaluation}\label{3}
Table II reports the hardware costs and Mean Absolute Error (MAE) for various multipliers for B=8-bit. Our proposed multiplier achieves 32.2\%, 42.8\%, and 51.8\% lower MAE compared to uMUL\cite{ugemm}, Jenson \cite{Jenson2016}, and Gaines \cite{Gaines1969}, respectively. In addition, our proposed multiplier achieves 10.6$\times 10^4$ better area-energy-latency product compared to the best prior work uMUL\cite{ugemm}. Moreover, we also show in Fig. \ref{stochasticmultiplier}(b) that the absolute error in the multiplication results from our multiplier is less dependent on the normalized difference of input operands ($|X_b-Y_b|/N$). This implies that our multiplier can provide stable accuracy irrespective of the input operand values, which is a desirable quality to have in multipliers used in GEMM accelerators.       

\section{Conclusions}
We presented a novel stochastic multiplier that generates stochastic bit-streams with reduced lengths and deterministic bit-position correlations in a bit-parallel manner, thereby simultaneously minimizing the latency, energy, and errors.

\bibliographystyle{IEEEtran}
\bibliography{references}

\end{document}